\documentclass[runningheads]{llncs}
\usepackage{graphicx}
\usepackage{amsmath,amssymb} 
\usepackage{color}

\begin{document}
\pagestyle{headings}
\mainmatter

\def\ACCV20SubNumber{199}  

\title{Utilizing Transfer Learning and a Customized Loss Function for Optic Disc Segmentation from Retinal Images} 
\titlerunning{Disc Segtmentation}
%
\author{Abdullah Sarhan\inst{1*} \and
Ali Al-Khaz’Aly\inst{1} \and
Adam Gorner\inst{2} \and Andrew Swift \inst{2} \and Jon Rokne \inst{1}\and Reda Alhajj \inst{1,4}\and Andrew Crichton \inst{3}}
\authorrunning{A. Sarhan et al.}
%
\institute{Department of Computer Science, University of Calgary, Calgary, AB, Canada \and
Cumming School of Medicine, University of Calgary, Canada \and Department of Ophthalmology and Visual Sciences, University of Calgary, Canada
\and Department of Computer Engineering, Istanbul Medipol University, Istanbul, Turkey\\
\email{*asarhan@ucalgary.ca}}

\maketitle

\begin{abstract}
Accurate segmentation of the optic disc from a retinal image is vital to extracting retinal features that may be highly correlated with retinal conditions such as glaucoma. In this paper, we propose a deep-learning based approach capable of segmenting the optic disc given a high-precision retinal fundus image. Our approach utilizes a UNET-based model with a VGG16 encoder trained on the ImageNet dataset. This study can be distinguished from other studies in the customization made for the VGG16 model, the diversity of the datasets adopted, the duration of disc segmentation, the loss function utilized, and the number of parameters required to train our model. Our approach was tested on seven publicly available datasets augmented by a dataset from a private clinic that was annotated by two Doctors of Optometry through a web portal built for this purpose. We achieved an accuracy of 99.78\% and a Dice coefficient of 94.73\% for a disc segmentation from a retinal image in 0.03 seconds. The results obtained from comprehensive experiments demonstrate the robustness of our approach to disc segmentation of retinal images obtained from different sources.

\end{abstract}

\section{Introduction}
Sight is one of the most important senses for humans, allowing us to visualize and explore our surroundings. Over the years, several degenerative ocular conditions affecting sight have been identified such as glaucoma and diabetic retinopathy. These conditions can threaten our precious sense of sight by causing irreversible visual-field loss \cite{sarhan2019glaucoma}. Glaucoma is the world’s second most prominent cause of irreversible vision loss after cataracts, accounting for 12\% of annual cases of blindness worldwide \cite{fu2017segmentation}. According to one estimate, around 80 million people are currently affected by glaucoma, and around 112 million will be affected by 2024. Approximately 80\% of patients do not know they have glaucoma until advanced vision loss occurs \cite{sarhan2019glaucoma,tham2014global}. 

The optic disc is one of the main anatomical structures in the eye which must be monitored and evaluated for progression when glaucoma is suspected \cite{sarhan2019glaucoma}. Changes within the optic disc, such as the displacement of vessels or enlargement of the optic cup to optic disc ratio can be used to help determine if glaucoma is present and if there is progression of the disease \cite{issac2015adaptive}. These changes occur because of an irreversible decrease in the number of nerve fibres, glial cells and blood vessels.

\begin{figure}[htb!]
\centering
\includegraphics[width=10cm,keepaspectratio]{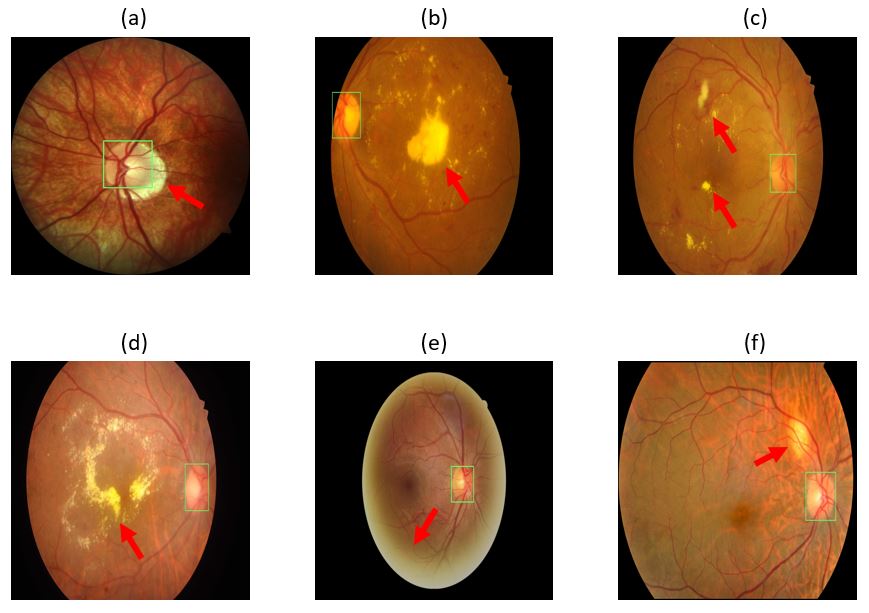}
\caption{Images showing various bright regions that can be observed in retinal images. The green box shows the location of the disc and the red arrows indicate other bright regions that may mislead some approaches when segmenting the disc.}
\label{fig:challangedImages}
\end{figure}

Several methods have been proposed for disc segmentation. These can be categorized as follows: morphological approaches \cite{panda2017robust}, template based matching approaches \cite{issac2015adaptive,sun2015optic}, adaptive-thresholding based approaches \cite{de2014application}, and pixel-classification based approaches \cite{zilly2017glaucoma}. Approaches related to the first three categories mainly fail in the presence of bright objects similar to the ones shown in Fig. \ref{fig:challangedImages} \cite{sarhan2020approaches}. The red arrows in Fig. \ref{fig:challangedImages} indicate some of the bright regions that can be encountered in retinal images that may affect disc segmentation.

With the rise of deep learning comes the potential for achieving high performance when segmenting a retinal image. Researchers have been working to develop models that place each pixel of a retinal image into a specific class during semantic segmentation. However, the performances of these approaches tend to decrease when new datasets emerge with different disc appearances or images with different resolutions. For instance, the deep-learning model M-Net, proposed by \cite{fu2018joint} performs well on the ORIGA dataset \cite{zhang2010origa}, but not on other datasets (e.g. the DRISHTI-GS dataset) \cite{sivaswamy2014drishti} as reported by \cite{wang2019patch}. One cause might be related to improper handling of the variance between classes when training. This is because the optic disc class may comprise between 2\%-10\% of an image depending on the angle and resolution of the captured image, whilst the background class would take up the rest of the image. This causes some models to converge toward the background and miss key details related to the disc.

In this paper, we propose a deep-learning approach to disc segmentation from retinal images using the UNET architecture to build the model and the VGG16 convolutional model as our encoder. Given the challenges related to having insufficient annotated disc datasets for deep learning, we adopted the idea of using transfer learning (TL) and image augmentation (IA). Instead of using random weights to initialize our model, we use weights trained on millions of images for semantic segmentation from the Imagenet dataset, which we then fine-tune to match the object we wanted to segment, which in our case is the optic disc. To handle the issue of imbalanced classes, we use a customized loss function that allows the loss function tend to penalize more when the wrong classification is made for a pixel related to a disc than that of background. The proposed model takes the whole retinal image as input and then segments the optic disc in a short time (milliseconds). To prove the robustness of our proposed approach in segmenting the disc with various sizes, angles, and orientations, we tested the approach on seven publicly available datasets and one private dataset that we formed \footnote{ https://github.com/AbdullahSarhan/ACCVDiscSegmentation}. 

Our contributions can be listed as follows: 
\begin{enumerate}
    \item We proposed a UNET based deep learning model for disc segmentation that uses VGG16 as the encoder.
    \item We demonstrated the effectiveness of using TL and IA for limited data.
    \item We handled the issue of imbalanced image classes which may lead to inaccurate results by adopting a weighted loss function.
    \item We contributed a new retinal image dataset for disc segmentation (ORDS). 
    \item We developed an online portal that can be used for annotating disc by multiple contributors.
\end{enumerate}

\section{Related Work}
\label{relatedWork}
Two types of approaches have been developed for disc segmentation: those that locate the optic-disc center but do not segment the disc and those that both locate the disc region and then segment the disc. In this section, we cover approaches that aim to segment the disc rather than just locating it.  

Earlier work entails the development of hand-crafted features that rely mainly on the shape of the disc and the intensity of pixels \cite{issac2015adaptive,sun2015optic,mohamed2019automated}. However, the performance of these hand-crafted approaches is easily affected by the presence of pathological regions and images with different resolutions (Fig. \ref{fig:challangedImages}). Recently, advancements made in the field of deep learning have opened the door to using deep learning based models in the field of medical-image analysis. Such approaches exhibit superior performance over the hand-crafted ones \cite{sarhan2019glaucoma,shen2017deep}.

Several approaches have been developed for segmenting the optic disc with deep learning: e.g., using an edge-attention guidance network to perform proper edge detection when segmenting the disc \cite{zhang2019net}, using disc-region localization and then disc segmentation via a pyramidal multi-label network \cite{yin2019pm}, using entropy-driven adversarial-learning models \cite{wang2019boundary}, using residual UNet based models \cite{baid2019detection}, and using generative adversarial networks combined with VGG16 and transfer learning \cite{jiang2019optic}.


In \cite{zilly2017glaucoma}, researchers used an ensemble-learning based convolutional neural-network model to segment the optic disc by first localizing the disc region. Entropy was used to select informative points; then the graph-cut algorithm was used to obtain the final segmentation of the disc. The researchers tested their approach only on the Drishti-GS \cite{sivaswamy2014drishti}, and RimOnev3\cite{pena2015estimation}, datasets. However, they used only 50 images from the Drishti-GS dataset with 40 for training and 10 for testing, even though the Drishti-GS dataset contains 101 images with 50 for training and 51 for testing. 


The use of transfer learning when working with deep learning to analyze medical images has been adopted by various studies \cite{shin2016deep,pan2009survey,karri2017transfer}. In the study performed by \cite{jiang2019optic} researchers adopted transfer learning to train their encoder for segmenting the disc when given the whole image without the need for cropping. They used the PASC AL VOC 2012 pretrained weights \cite{everingham2010pascal}. They used only the Drishti-GS dataset in both training and testing their model. Moreover, the number of training parameters used by their model, $30.85 * 10^6$, is double that of our approach. In \cite{wang2019patch} transfer learning for disc segmentation was also used. They started by cropping the disc region using the UNET model developed in \cite{ronneberger2015u} initializing their encoder using the weights of the MobileNetV2 \cite{sandler2018mobilenetv2} trained on the ImageNet dataset \cite{russakovsky2015imagenet}.

The majority of the developed approaches tend to first locate the disc region and then feed this region into their model to avoid bright regions like the one shown in Fig. \ref{fig:challangedImages}. For proper segmentation, such approaches are highly dependent on successful localization of this disc region. Moreover, these approaches tend to handle the issue of imbalanced classes, which makes their approach perform differently when new images with different resolutions emerge. In this paper, we show a model with an encoder that uses transfer learning, proper data augmentation, and a customized loss function can segment the optic disc with high precision, giving results comparable to the above-mentioned approaches. In this study, we do not localize the disc prior to performing the segmentation and instead feed the whole image to our model, rather than a specific region of the image.



\section{Proposed Method}
Our goal is to segment the optic disc given a retinal image. To achieve this, we propose a deep learning model with the same architecture as the UNET model \cite{ronneberger2015u}. A pixel matrix $I$ is associated with each retinal image indicating the pixels either belong to the disc and the background. If $I_{xy}$ represents a pixel at location $(x,y)$ in the retinal image, this pixel will have a value of 1 if it belongs to the disc and 0 if it is a background pixel. The model will use these labeled images and the actual retinal image to produce a new image with the same dimensions, where each pixel has a probability between 0 and 1 inclusive, thus indicating whether this pixel belongs to the disc or not. The closer the value is to 1, the higher the model's confidence is that it belongs to a disc. In this section, we describe the model adopted in this study.

\subsection{Network Architecture}
\begin{figure}[htb!]
\centering
\includegraphics[width=10cm,height=10cm,keepaspectratio]{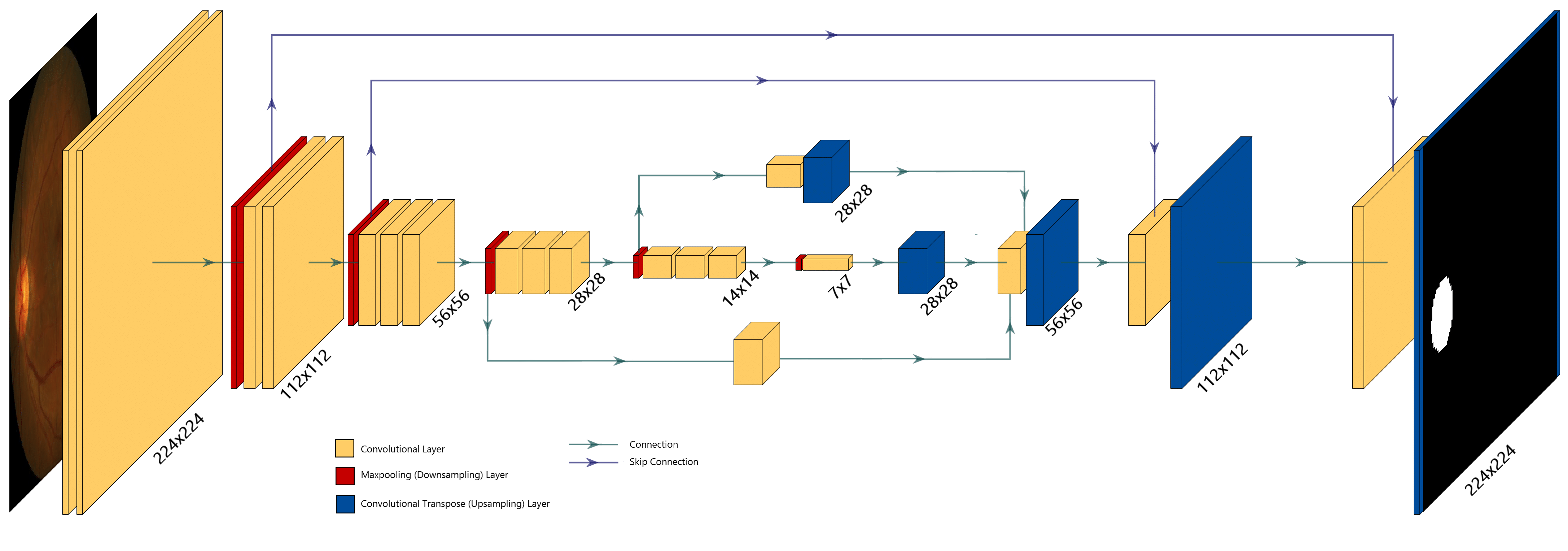}
\caption{Architecture of the Customized VGG-16 model Adopted in this Study.}
\label{fig:VGG16ModelArchitecture}
\end{figure}

Instead of creating a new architecture, we adopted the U-Net architecture which consists of an encoder and a decoder. The encoder is responsible for down-sampling the image, and the decoder is responsible for up-sampling the image to provide the final output. In our case, we used the VGG16\cite{simonyan2014very} model as the encoder and built the decoder by using a series of skip connections, convolutional, up-sampling, and activation layers, as shown in Fig. \ref{fig:VGG16ModelArchitecture}. 

The original VGG16 model with a down-sampling factor of 32 is customized so that it could be used for semantic segmentation. It contains five down-sampling layers followed by two densely connected layers and a softmax layer for prediction. We removed the two densely connected layers at the end of the original model and replaced them with a single convolutional layer found in the center of our model, as shown in Fig. \ref{fig:VGG16ModelArchitecture}. Doing so reduced the number of parameters used to train the model from 134,327,060 to 16,882,452. Fixing this bottleneck significantly cut down the time and computational power required to train the model without causing any observable changes to the model’s predictions.
We also removed the softmax layer and added all of the upsampling and convolutional layers seen on the right half of the model as is needed in image segmentation to regenerate the original image shape, finally our last layer is a sigmoid activation layer which predicts on the feature matrix. The 5 upsampling layers achieve an upsampling factor of 32, allowing the output images to have the same shape as input images, counteracting the data reshaping effects caused by down-sampling layers. The feature map for each convolutional layer is the ReLU \cite{kingma2014adam} activation method, which applies Eq. \ref{equ:relU} to each parameter coming out of the layer, thereby removing all negative pixel values.
\begin{equation}
\label{equ:relU}
f(x)={max(0,x)}
\end{equation}
Throughout the architecture, there are several instances where we use skip connections. In Fig. \ref{fig:VGG16ModelArchitecture}, the first and second maxpooling layers utilize a skip connection to convolutional layers further down in the pipeline of the model. Whilst the third and fourth maxpooling layers are connected directly to a convolutional layer which attaches either directly or indirectly to the later layers. This was used to shorten the distance between the earlier and later layers. Short connections from early to later layers are useful in preserving high-level information about the positioning of the disc. This is opposed to the low-level pixel-based information which is transferred across the long pipeline of the architecture in a combination of convolutional and maxpooling/upsampling layers. High-level information tends to be lost as the image gets down-sampled and the shape and structure of the image is changed. Therefore, we maintain this information by using connections to earlier layers in the model. Better results were observed when using instead of not using these connections.

\subsection{Transfer Learning}
To handle the challenge faced in the field of medical imaging of not having enough datasets or large enough datasets to train a deep-learning model, we used an approach referred to as transfer learning. As discussed in Section \ref{relatedWork}, such approaches can alleviate the issues caused by insufficient training data by using weights generated by training on millions of images \cite{pan2009survey}. In our study, we adopted the weights generated when training the VGG16 model on the ImageNet dataset \cite{russakovsky2015imagenet} which contains around 14 million labeled images. We thus provided a diverse set of images that the model had been exposed to.\\

By using transfer learning, we could reduce the problem of over-fitting caused when training on limited images and improve the overall performance of the model. Using the ImageNet weights, we initialized the weights of the encoder network component, and other layers were randomly initialized using a Gaussian distribution. We then trained our model using a mini-batch gradient to tune the weights of the whole network. When training, we realized when using transfer learning that the model converged faster than without transfer learning.

In addition to transfer learning, we applied random augmentation to each image. by randomly applying any of the following: horizontal shifting, vertical shifting, rotation within a range of 360 degrees, horizontal flipping, vertical flipping, or any combination of the above. We tested the evaluation effectiveness of data augmentation with and without transfer learning.

\subsection{Loss Function}
During the training of the network, we decided whether a model had improved on the value returned from the loss function by running it on validation data. Initially, we adopted the binary cross-entropy function (BCE), as shown in Eq. \ref{equ:BCE} where $N$ is the number of all pixels, $y_i$ is the label of that pixel (0 for background and 1 for the disc), $p(y_i)$ is the predicted probability that the pixel belongs to the disc and $p(y_i)$ is the predicted probability of being a background pixel. Note that the BCE can penalize both false positives and false negatives when working with foreground and background classification.
\begin{equation}
\label{equ:BCE}
BCE=- \frac{1}{N}\sum_{i=1}^N y_i.\log (p(y_i)) +(1-y_i).\log(1-(p(y_i)))
\end{equation}
For any given retinal image, the disc will be only occupy a small region of the image (usually 2-10\%), with the large majority of the image being background, i.e. 90\% or more. Using this loss function alone would therefore not be sufficient for a precise disc segmentation output. This is because the BCE will be biased toward the background and hence, the disc will not be properly segmented. Thus, it may give an accuracy of 90\%, which may be misleading. To bypass this issue, we decided also to use the Jaccard distance. The Jaccard distance measures how dissimilar two sets of data are. The Jaccard loss function is defined as:
\begin{equation}
\label{equ:JLoss}
L_{j}=1- \frac{|Y_d \cap \hat{Y_d}|}{|Y_d \cup \hat{Y_d}|}
= 1- \frac{\sum_{d\in Y_d} (1 \land \hat{y_d})}{|Y_d| + \sum_{b\in Y_b} (0 \lor \hat{y_b})}
\end{equation}
where $Y_d$ and $Y_b$ represent the ground truth of the disc and background respectively. $\hat{Y_d}$ and $\hat{Y_b}$ represent the predicted disc and background pixels. $|Y_d|$ $|\hat{Y_b}|$ represents the cardinality of the disc $Y_d$ and background $\hat{Y}_b$ respectively with $\hat{y_d} \in \hat{Y_d}$
 and $\hat{y_b} \in \hat{Y_b}$. Since $\hat{Y_d}$ and $\hat{Y_b}$ are both probabilities, and their value will always be between $0$ and $1$, we can approximate this loss function as shown in Eq. \ref{equ:JApproximated} and the model will then be updated by Eq. \ref{equ:JDerived} where j represents the the $j$th pixel of the input image and $\hat{y_j}$ represents the predicted value for that pixel.
\begin{equation}
\label{equ:JApproximated}
\tilde{L_j} = 1- \frac{\sum_{d\in Y_d} min(1, \hat{y_d})}{|Y_d| + \sum_{b\in Y_b} max(0,\hat{y_b})} \\
= 1- \frac{\sum_{d\in Y_d} \hat{y_d}}{|Y_d| + \sum_{b\in Y_b} \hat{y_b}}
\end{equation}
\begin{equation}
\label{equ:JDerived}
{L_j}{y_i} 
\begin{cases}-\frac{1}{|Y_d| + \sum_{b\in Y_b} \hat{y_b}} &  \hspace{0.5cm} for  \hspace{0.5cm} i \in Y_d\\
\\-\frac{\sum_{d\in Y_d} \hat{y_d}}{|Y_d| + \sum_{b\in Y_b} \hat{y_b}} & \hspace{0.5cm} for \hspace{0.5cm} i \in Y_b\end{cases}
\end{equation}
Given the Jaccard loss function, we are able to balance the emphasis the model gives to each of the classes: namely, the disc class and background class. In this, we combine BCE with Jaccard to optimize the results. We realize that, when both are combined, the model can converge faster than it can when using only the Jaccard while still achieving better results than BCE or Jaccard alone. Hence, our final loss function is:
\begin{equation}
\label{equ:FinalLoss}
Loss=BCE+L_j
\end{equation}

\subsection{Implementation Details}

To implement this model we used a windows machine with a NVIDIA GeForce 2060 RTX with 6 GB dedicated GDDR6 memory and 8GB of shared random access memory which the GPU is free to use as necessary. We used the Python language to implement the proposed approach using Keras with TensorFlow back-end.

Training was performed using the NAdam optimizer \cite{dozat2016incorporating} function with learning rate set to 0.0001, $\beta_1$ = 0.9, $\beta_2$ = 0.999, $\epsilon= 10^{-8}$, and batch size of 4 images. During training, three callbacks were used. First, the model checkpoints would save the model whenever a smaller value was returned on validation data from the custom loss function when comparing to the value at the last checkpoint. Secondly, the learning rate was reduced by a factor of 0.5 whenever 25 epochs passed without any improvement in the validation loss values. Finally, the training was stopped if 100 epochs passed without any improvement.

\section{ORDS Dataset}
Datasets obtained from different resources had to be used in order to evaluate the reliability and applicability of the proposed method. One of the issues faced when working with disc segmentation is the lack of diverse datasets. To augment the available data sets we decided to contribute a new dataset obtained from a private clinic, annotated by two experts in this field. In this section, we discuss the new data collected.

\begin{figure*}[htb!]
\centering
\includegraphics[width=10cm,height=10cm,keepaspectratio]{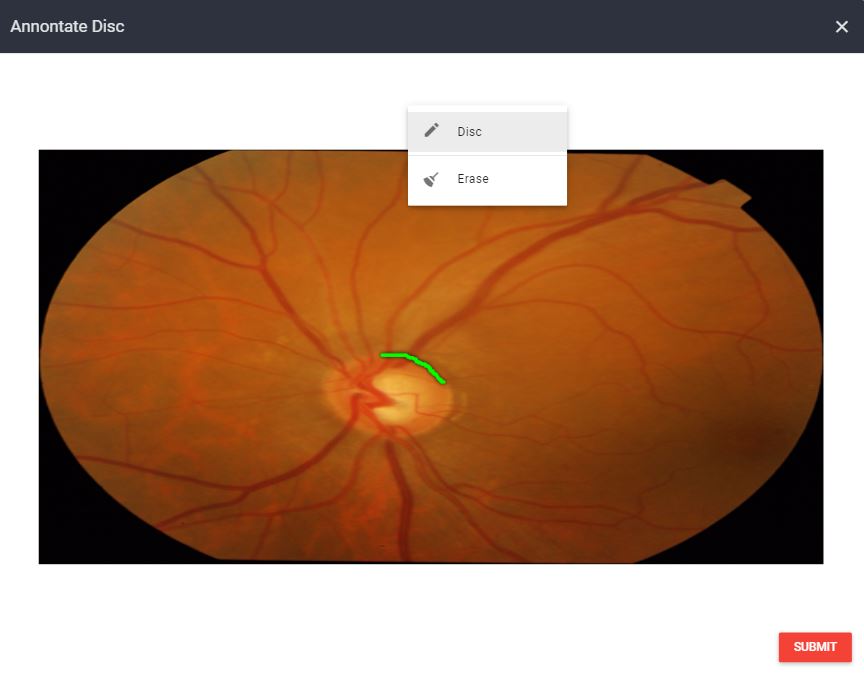}
\caption{Web portal showing images assigned to the Optometrists along with the tracing form utilized for disc tracing.}
\label{fig:tracing}
\end{figure*}

The ORDS dataset, our new dataset \footnote{ https://github.com/AbdullahSarhan/ACCVDiscSegmentation}, was obtained from a private clinic in Calgary, and the disc was annotated by two Doctors of Optometry. We built a customized web portal to help optometrists trace the disc\footnote{ The portal is now publicly available. Link and login credentials can be provided upon request}. Each optometrist was assigned a username and password to log into the portal and view the assigned images. Upon successful login, a user can navigate to the tracing page and start tracing, as shown in Fig.\ref{fig:tracing}. Both optometrists traced the same set of images; and hence each image received two annotations for the disc. In total, 135 images were annotated.
On the tracing page, a list of images is presented; the user can click on any image, and a pop-up dialogue will appear. Once the pop-up model appears, the user can start tracing; an erase option is presented should the user wish to erase any of the tracing. Once the tracing is done, the user can click on the submit button, which will allow the storage of tracing information on a dedicated server. Users have the option either to trace the whole disc at once or in steps. Upon successful submission of the tracing, the traced image will be eliminated from the list of images on the tracing page.

\section{Experimental Results}
We evaluated our methods on eight different datasets which allows us to evaluate our approach when discs with different sizes, orientations, and resolutions are fed to our model. In this section, we discuss the datasets adopted, experiments conducted, and compare the performance of our model with other approaches.

\subsection{Datasets}

\begin{figure}[htb!]
\centering
\includegraphics[width=\linewidth,keepaspectratio]{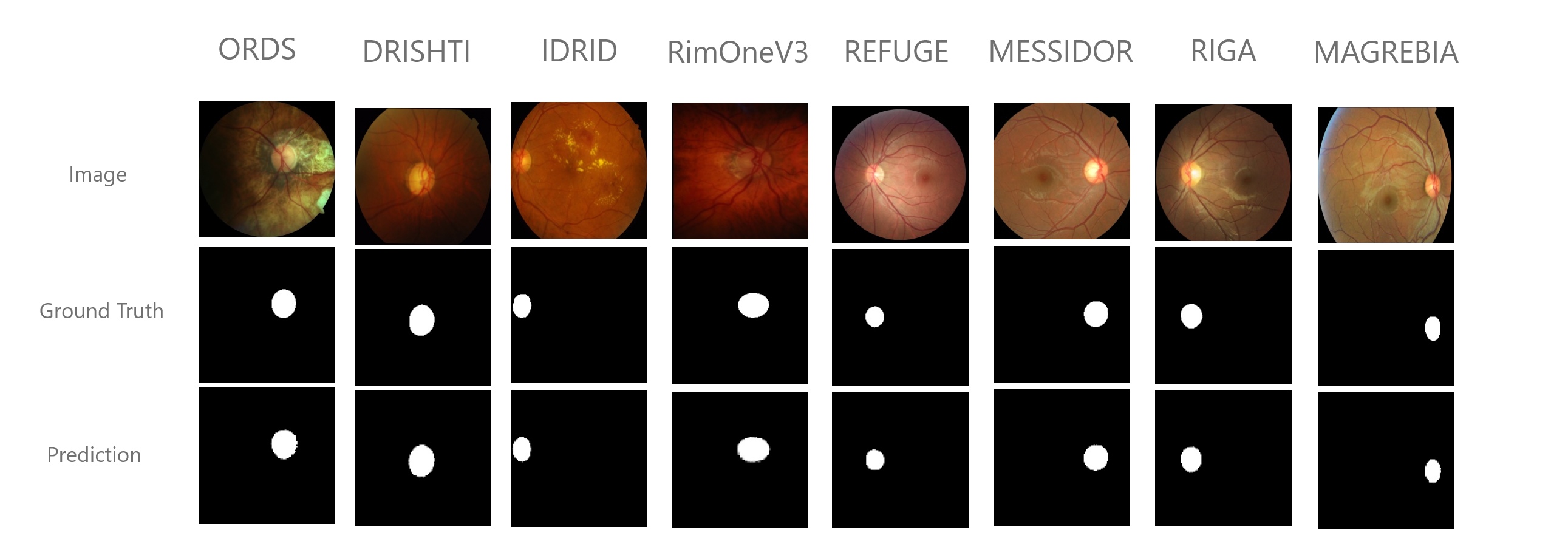}
\caption{A sample image from each dataset used in our study. The first row shows the actual retinal image while the second and third shows the related ground truth and prediction made by our model respectively. The name of the dataset to which each image belongs is written at the top of its column.}
\label{fig:OutputModel}
\end{figure}

To verify the robustness of our method, we tested our approach on seven publicly available datasets and our dataset. Table \ref{table:Of} provides an overview of these datasets along with the machines used to capture these images, including our new dataset. These datasets contain information regarding multiple retinal conditions: namely, glaucoma and diabetic retinopathy. Moreover, retinal images that belong to these datasets were acquired at different angles and resolutions, as can be seen in Fig. \ref{fig:OutputModel}. For datasets that contained multiple annotations, including our new dataset which had two expert tracings, we used the average of the tracings when training and evaluating our model, which is the common technique used in such scenarios \cite{pena2015estimation,sivaswamy2014drishti}.
\begin{table}[htb!]
\center
\caption {\label{table:Of} Dataset properties and machines used to capture their images. }
\begin{tabular}{lcccc}
\hline
Dataset                                               & \multicolumn{2}{c}{Images} & Dimensions & Machine                      \\ \cline{2-3}
                                                      & Train        & Test        &            & \multicolumn{1}{l}{}         \\ \hline
Drishti-GS \cite{sivaswamy2014drishti}  & 50           & 51          & 2049*1757  & -                            \\ \hline
Refuge \cite{orlando2020refuge}      & 400          & 400         & 2124*2056  & Zeiss Visucam 500            \\ \hline
IDRID \cite{porwal2020idrid}         & 54           & 27          & 4288*2848  & Kowa VX-10 alpha             \\ \hline
Rim\_r3 \cite{pena2015estimation}    & 128          & 31          & 1072*712   & Nidek AFC-210                \\ \hline
BinRushed \cite{almazroa2018retinal} & 147          & 35          & 2376*1584  & Canon CR2 non-mydriatic      \\ \hline
Magrebia \cite{almazroa2018retinal}  & 52           & 11          & 2743*1936  & Topcon TRC 50DX mydriatic    \\ \hline
Messidor \cite{almazroa2018retinal}  & 365          & 92          & 2240*1488  & Topcon TRC NW6 non-mydriatic \\ \hline
ORDS                                & 110          & 25          & 1444*1444  & Zeiss, Visucam 200           \\ \hline
\end{tabular}
\end{table}
In total 1,442 images were used for training and 705 were used for testing. To test our model, we had to have the data split into training and testing portions. The model could only see the training images, and we checked the performance by evaluating the model's predictions on the test images and comparing it to labels. Doing so makes it fair to compare our model with other approaches, as they would test their approach on the same test images we are using. However, not all datasets are split in this manner. For some datasets, we had to do the splitting with 75\% of the dataset used for training and 25\% for testing, selection done randomly. We did this split for the Messidor, ORDS, BinRushed, Magrebia, and RimOneV3 datasets. Note that the annotation used for Messidor is different from that used by other approaches,(e.g. \cite{thakur2019optic}) as the annotation used by such studies is not available anymore. We used the one provided by \cite{almazroa2018retinal}. The testing images for all datasets are provided in our supplementary material so that other researchers can make fair comparisons to ours, and thereby standardize the images these comparisons are made on\footnote{https://github.com/AbdullahSarhan/ACCVDiscSegmentation}. Fig. \ref{fig:OutputModel} shows the performance of our model on a test image from each dataset, each dataset being different in terms of angle and resolution.

Doing this allows other researchers to compare their approaches by standardizing the set of test images without using the leave-one-out strategy  ,\cite{wang2019blood}, which would be time-consuming due to the number of training experiments that must be conducted for each dataset. For instance, if we have a dataset with 200 images then we need to train our model 200 times each time using 199 images and test on the excluded image; we would have to train 200 models and average the test results of them. 
Images found across different datasets and even within a single dataset can be extremely inconsistent in shape, size of optic disc region, and pixel values. Therefore, a general rule is applied to preprocess all images before they are passed to the model. They are first resized to 224*224 pixels, normalized so that all pixel values are within the range (0,1), and finally, undergo binary thresholding for disc ground-truth images.

\subsection{Evaluation Methods}
We used four evaluation methods to evaluate and compare our approach: namely, accuracy (Acc):$\frac{TP+TN}{ TP+FP+TN+FN}$, dice coefficient (DC):$2* \frac{Area(A \cap B)}{ Area(A)+ Area(B)}$, sensitivity (Sen): $\frac{TP}{TP+FN}$, and intersection over union (IoU):$\frac{Area(A \cap B)}{ Area(A) \cup Area(B)}$ . Moreover, we also show the time required by our approach to segment the disc and compare it with information obtained by other approaches (when applicable).








\subsection{Effectiveness of TL and IA}
To test the impact of using transfer learning (TL) and image augmentation (IA) when training our model we conducted a series of experiments and then evaluated the model obtained using the test images for all datasets. In this section we show the overall performance without showing performance on each dataset. Note that in all these experiments we used the loss function defined in Eq. \ref{equ:FinalLoss}.

We first checked the performance of the model without transfer learning by randomly initializing weights using Gaussian distribution, which needed 128 epochs to finish training. Then we did an experiment using data augmentation also without transfer learning and this needed 141 epochs. The third and fourth experiment using TL but with and without IA and they needed 184 and 207 epochs to finish training respectively. The evaluation results for each of these experiments are shown in Table \ref{table:TL}. The results obtained show that using TL and IA together achieve the best results especially for DC and IoU values, which really reflect how precisely the disc is segmented along with it being slightly faster than the other models. 
\begin{table}[htb!]
\center
\caption {\label{table:TL} Performance comparison of proposed method with and without using transfer learning (TL) and/or image augmentation (IA). }
\begin{tabular}{lccccc}
\hline
Experiment      & Acc   & DC    & Sen   & IoU &Time(s)  \\ \hline
No TL and No IA & 99.68 & 92.41 & 97.01 & 86.41 &0.0317\\ \hline
IA with no TL     & 99.74 & 93.80  & 96.18 & 88.59& 0.0366\\ \hline
TL with No IA   & 99.72 & 93.41 & 97.13 & 87.94& 0.0308\\ \hline
TL with IA      & 99.78 & 94.73 & 96.26 & 90.13 &0.0306\\ \hline
\end{tabular}
\end{table}

\subsection{Effectiveness of Loss Functions}
A well known loss function for binary classification is the binary cross entropy loss function. This loss function works great when the classes in the image are balanced. However in our case, the object we are trying to segment represents 10\% or less of the total image area of the image. Hence, we decided to use the Jaccard distance approach as noted earlier.

We conducted three experiments to test which configuration would achieve the best results. First, we trained our model using the BCE loss function alone, which is a built-in loss function in the keras library, second, we trained using only the Jaccard loss function and finally, we trained using a combination of both loss functions. The results obtained are shown in Table \ref{table:lossF}. We realized there is slight improvement in performance when we combine both loss functions compared to using either one of them alone. We also realized that using Jaccard alone achieved better results than BCE but it took 516 epochs to finish training compared to 210 epochs when using BCE alone and 374 epochs when combining both. Note that in all these experiments we used TL and IA.
\begin{table}[htb!]
\center
\caption {\label{table:lossF} Performance of the model across different loss functions.}

\begin{tabular}{lccccc}
\hline
Loss Function         & Acc   & DC    & Sen   & IoU   & Time(s) \\ \hline
BCE                   & 99.75 & 94.01 & 94.89 & 88.82 & 0.0358  \\ \hline
Jaccard Distance      & 99.76 & 94.03 & 95.72 & 88.85 & 0.0329  \\ \hline
Jaccard Distance+BCE & 99.78 & 94.73 & 96.26 & 90.13 & 0.0306  \\ \hline
\end{tabular}
\end{table}

\subsection{Comparing with Other Approaches}
To evaluate our proposed method we compared with approaches which were tested on some of the same datasets we used, as shown in Table \ref{table:OP}. Unfortunately, these approaches did not evaluate using all available datasets and hence when comparing we split our results per dataset to be able to do a fair comparison. We achieved an overall average accuracy of 99.78\%, DC of 94.73\%, Sensitivity of 96.26\% and IoU of 90.13\%. Our approach outperformed other approaches tested on some of the online publicly available datasets as shown in Table \ref{table:OP} except two approaches for some of the dataset they used. Further, we achieved a prediction time that is the best among the current state of the art approaches with average segmentation time is 0.03s.
\begin{table}[htb!]
\center
\caption {\label{table:OP} Performance comparison of proposed method on optic disc segmentation. }

\begin{tabular}{llccccc}
\hline
Method                                      & Dataset             & \multicolumn{4}{c}{performance metrics}                                                                           & \multicolumn{1}{l}{Time(s)} \\ \cline{3-6}
                                            &                     & \multicolumn{1}{l}{Acc} & \multicolumn{1}{l}{DC} & \multicolumn{1}{l}{Sen} & \multicolumn{1}{l}{IoU} & \multicolumn{1}{l}{}         \\ \hline
\textbf{Wang et al. \cite{wang2019boundary}}       & \textbf{RimOnev3} & -                            & 89.80                   & -                               & -                       & -                            \\
                                            & \textbf{Drishti-GS} & -                            & 96.40                   & -                               & -                       & -                            \\ \hline
\textbf{PM-Net \cite{yin2019pm}}                             & \textbf{Refuge}    & 97.90                         & -                      & -                               & -                       & -                            \\ \hline
\textbf{ET-Net \cite{zhang2019net}}            & \textbf{Refuge}    & -                            & 92.29                  & -                               & 86.70                    & -                            \\
                                            & \textbf{Drishti-GS} & -                            & 93.14                  & -                               & 87.90                    & -                            \\ \hline
\textbf{Thakur et al. \cite{thakur2019optic}}      & \textbf{RimOneV3} & 94.84                        & 93.00                     & -                               & -                       & 38.66                        \\
                                            & \textbf{Drishti-GS} & 93.23                        & 92.00                     & -                               & -                       & -                            \\ \hline
\textbf{GAN-VGG16 \cite{jiang2019optic}}                          & \textbf{Drishti-GS} & -                            & \textbf{97.10}          & -                               & -                       & 1                            \\ \hline
\textbf{ResUNet \cite{baid2019detection}}                            & \textbf{IDRID}      & -                            & 86.50                   & -                               & -                       & -                          \\ \hline
\textbf{pOSAL \cite{wang2019patch}}          
& \textbf{Refuge}     & -& \textbf{96.00}    & -& -& -\\
& \textbf{Drishti-GS} & -& 96.50& -& -& -\\
& \textbf{RimOneV3}  & -& 86.50& -& -& -\\ \hline
Proposed Approach& \textbf{Drishti-GS} & \textbf{99.79} & 96.50  & \textbf{97.54}  & \textbf{93.18} & \textbf{0.03} \\
                                                                                   & \textbf{IDRID}      & \textbf{99.80} & \textbf{95.39} & \textbf{96.94} & \textbf{91.30}  & \textbf{0.12} \\
                                                                                   & \textbf{RimOneV3}  & \textbf{99.50}  & \textbf{94.91}    & \textbf{96.11} & \textbf{90.44} & \textbf{0.03} \\
                                                                                   & \textbf{Refuge}     & \textbf{99.80} & 94.09                           & \textbf{95.77} & \textbf{89.00} & \textbf{0.02} \\
                                                                                   & \textbf{BinRushed}  & \textbf{99.82} & \textbf{95.57} & \textbf{96.97} & \textbf{91.53} & \textbf{0.03} \\
                                                                                   & \textbf{Magrebia}   & \textbf{99.80} & \textbf{96.18} & \textbf{95.58} & \textbf{92.68} & \textbf{0.04} \\
                                                                                   & \textbf{Messidor}   & \textbf{99.83} & \textbf{96.16}    & \textbf{97.18} & \textbf{92.62} & \textbf{0.03} \\
                                                                                   & \textbf{ORDS}       & \textbf{99.50} & \textbf{93.58} & \textbf{96.83} & \textbf{88.25} & \textbf{0.03} \\ \cline{2-7} 

               \hline
\end{tabular}
\end{table}

For the Refuge dataset we achieved better results than the ones reported by \cite{yin2019pm} and \cite{zhang2019net} yet we achieved slightly lower than the values reported by \cite{wang2019patch} whom reported achieving 96\% where we achieved 94.09\%. However, we achieved better than them in the RimOneV3 dataset and Drishti-GS. Note that they first localize a region of interest and then segment the disc whereas in our case we directly segment the disc from the whole retinal image without first localizing the region the disc is located in. 

For the Drishti-GS dataset our model performed better than \cite{wang2019boundary,zhang2019net,thakur2019optic}, the same as \cite{wang2019patch}, and slightly lower than \cite{jiang2019optic}. However, in \cite{jiang2019optic} they only trained and tested their approach one the Refuge dataset, which is not enough to show how well their system work on images from multiple sources. Moreover, their model requires $30.85 * 10^6$ parameters which is almost double what our model requires. 

Our model achieved better results than the approaches mentioned above for the IDRID and RimOneV3 datasets. For the dataset provided by \cite{almazroa2018retinal} they are still a new dataset and up to our knowledge there is no study with published testing images that we can use to compare the performance of our model with. To ensure continuity of this research and allow researchers to be able to perform fair comparison we will publish all test images used to evaluate our model in our supporting material. We also publish both the training history log and our model which was tested on in Table \ref{table:OP}. In general our model demonstrated high performance segmenting the disc for images obtained from different resources with different angles of the disc and resolutions, including challenging ones as shown in Fig. \ref{fig:challangedImages} (check supplementary material for more images).


\subsection{Leave One Out Experiment}

Clinics may capture images with different resolutions and angles. To verify the robustness of our model on images that it was not trained on, that may have different characteristics than what it was trained on, we conducted 8 experiments where a model was trained on all datasets except for one which was used for evaluation. The results obtained for each dataset are showing in Table \ref{table:LO}. This table shows that for instance, when the model is trained on all datasets except for Refuge, it will achieve a DC value of 92.51\% which is slightly less than when using the cross training which is 94.09\%. The results seem consistent in that our model can effectively segment the disc, except for the RimOneV3 dataset. This is likely because this dataset only provides the images of the area surrounding the disc.
\begin{table}[htb!]
\center
\caption {\label{table:LO} Performance of the model when being trained on all datasets except for the one being evaluated on. }

\begin{tabular}{lccccc}
\hline
Dataset    & Acc   & DC    & Sen   & IoU   & Time(s) \\ \hline
Drishti-GS & 99.76 & 95.94 & 96.82 & 92.25 & 0.07    \\ \hline
IDRID      & 99.66 & 92.47 & 97.83 & 86.11 & 0.03    \\ \hline
RimOneV3  & 98.23 & 80.00 & 89.49 & 70.00 & 0.03    \\ \hline
Refuge     & 99.75 & 92.51 & 96.34 & 86.42 & 0.02    \\ \hline
BinRushed  & 99.74 & 95.01 & 92.67 & 90.54 & 0.03    \\ \hline
Magrebia   & 99.76 & 95.68 & 93.77 & 91.74 & 0.04    \\ \hline
Messidor   & 99.66 & 94.11 & 97.83 & 90.00 & 0.03    \\ \hline
ORDS       & 99.29 & 89.29 & 86.00 & 81.51 & 0.03    \\ \hline
\end{tabular}
\end{table}

\section{Conclusion and Future Work}
In this paper, we proposed a deep learning based approach for disc segmentation where we proved the effectiveness of transfer learning, image augmentation, and a customized loss function. Our approach achieved state of the art performance on disc segmentation when compared to other modern approaches. We also contribute a new dataset the can be used by researchers for improving disc segmentation. This will help researchers testing their approaches on images obtained from various sources with diverse data. Our new dataset was annotated by two doctors of optometry using an online portal we built for the annotation task. 

As for future work, we would like to expand our approach to include glaucoma detection by analyzing the disc region. Using the cup/disc alone is not always an indicator for glaucoma and hence we need to analyze the disc region and make an assessment. Moreover, we would like also to expand our portal to be used for educational and research purposes where people can share and annotate the datasets. Further, we would like to improve our dataset to include annotation for other anatomical objects in the retina such as peripapillary atrophy and exudates.



\bibliographystyle{splncs}
\bibliography{accv2020submission}

\end{document}